\documentstyle[12pt,psfig]{article}
\newcommand{\mysection}{\setcounter{equation}{0}\section}

\begin{document}

\pagestyle{myheadings}
\hfill{ITP-SB-93-80}
\vskip 1cm
\centerline{\large\bf Exclusive $W^+ + \gamma$ production in}
\centerline{\large\bf proton-antiproton collisions II: results}
\vskip 0.5cm
\centerline{\bf S. Mendoza and J. Smith}
\centerline{\small \sl Institute for Theoretical Physics,}
\centerline{\small \sl State University of New York at Stony Brook,}
\centerline{\small \sl Stony Brook, New York 11794-3840}
\vskip 1cm
\centerline{February 1994}
\vskip 0.5cm
\centerline{\bf Abstract}

We present results for total cross sections, single and double differential
distributions and correlations between pairs of outgoing particles in the
reactions $p + \bar p \rightarrow W^+ + \gamma$ and $p + \bar p \rightarrow W^+
+ \gamma + jet$ at $\sqrt{S}=1.8$ TeV. Order $\alpha_S$ QCD corrections and
leading logarithm photon bremsstrahlung contributions are included in the
$\overline{MS}$ mass factorization scheme for three experimental scenarios:
1) 2-body inclusive production of $W^+$ and $\gamma$, 2) exclusive production
of $W^+$, $\gamma$ and $1$ jet and 3) exclusive production of $W^+$ and
$\gamma$ with $0$ jet.
The latest CTEQ parton distribution functions, which fit the newly released
HERA data, are used in our analysis. The dependence of our results on the mass
factorization scale is used
to place error bars on our predictions for the single differential
distributions and correlations.

\newpage
\pagestyle{myheadings}
\mysection{\bf INTRODUCTION.}

In the first part of this study \cite{MS1} we presented the general formalism
for the computation of exclusive cross sections in $p +\bar p  \rightarrow W^+
+\gamma$ and $p +\bar p  \rightarrow W^+ +\gamma + jet$. We examined the order
$\alpha_S$ and
photon bremsstrahlung contributions to these cross sections in the
$\overline{MS}$ mass factorization scheme. The analytical
expressions obtained in that work were then combined with a Monte Carlo
integration routine, namely Vegas \cite{P}, to produce numerical results for
three different experimental scenarios:
1) 2-body inclusive production of $W^+$ and $\gamma$, 2) exclusive production
of $W^+$, $\gamma$ and $1$ jet and 3) exclusive production of $W^+$ and
$\gamma$ with $0$ jet.
The resulting programs have been run for $\sqrt{S}=1.8$ TeV in the
proton-antiproton center of mass system (Fermilab Tevatron). We include 4
massless flavours $u,d,s$ and $c$ in our partonic hard scattering processes and
set $\cos\theta_C = 0.95$ where $\theta_C$ is the Cabbibo angle. We use the one
loop strong running coupling constant with $\Lambda_{QCD}=0.139$ GeV. The $W$
boson is assumed to be an on shell physical particle with mass $M_W=80.2$ GeV.

We use the latest available CTEQ parton distribution functions \cite{BLMOTW} in
the $\overline{MS}$ mass factorization scheme (set CTEQ2M). These parton
distribution functions are in agreement with the latest CCFR NLO analysis of
the strange quark density and fit the newly released HERA data (for details see
\cite{BLMOTW}.)

The three experimental scenarios considered here are defined as follows (see
also the discussion in \cite{MS1}):\\

{\sc 1) $2$ body inclusive production of $W^+$ and $\gamma$.\\}

In this scenario (``$2$ body inclusive scenario'') the reaction $p + \bar p
\rightarrow W^+ + \gamma + X$ with $X=0$ or $1$ jet is considered. In other
words, one detects the outgoing $W^+$ boson and the photon but does not tag the
outgoing jet.
We therefore use the following kinematic restrictions on the outgoing
particles, in the proton-antiproton center of mass frame:
\begin{eqnarray}
\label{1.1}
|\cos\theta_{\gamma}|, |\cos\theta_{W}| &<& \cos(0.3\ {\rm rad}) \nonumber \\
Pt_{\gamma}, Pt_{W}  &>& 10.0\ {\rm GeV} \nonumber \\
R_{W,\gamma} &>& 0.2 \nonumber \\
( R_{jet,\gamma} < 0.2 )&\Longrightarrow& (s_{(jet,\gamma)} < 0.2) \nonumber \\
( R_{jet,W} < 0.2 )&\Longrightarrow& (s_{(jet,W)} < 0.2) \,,
\end{eqnarray}
where $\theta_i$ is the angle between the incoming proton axis and the axis of
the outgoing particle $i$ and $Pt_{i}$ is the transverse momentum of particle
$i$. $R_{i,j}$ is the cone size between a pair of outgoing particles:
$R_{i,j}=\sqrt{(\eta^{\ast}_{i,j})^2 + (\phi_{i,j})^2}$ with $\eta^{\ast}$ the
pseudorapidity
and $\phi$ the azimuthal angle; $s_{(jet,W)}=E_{jet}/E_W$ is the ``shadowing
ratio'' between the untagged
jet and the $W^+$ boson. The third condition in (\ref{1.1}) removes events
where the $W^+$ boson is too close to the photon. The last two conditions
remove events where the jet which is too close to the $W^+$ boson or photon is
at the same time of comparable energy so that it would ``shadow'' one of the
two tagged particles making it undetectable. This might be the case for
bremsstrahlung contributions with a small photon momentum fraction.

Several artificial parameters, namely $x_0,y_0$ and $v_0$ were introduced in
\cite{MS1} in order to control the numerical cancellation of singularities.
Results for physical quantities like total cross sections and differential
distributions and correlations do not depend on the choice of these parameters,
however, the requirement of stability and small numerical errors in the Monte
Carlo runs lead us to the choice $x_0 = (1+\rho(s))/2$, $y_0 = 1.0$ for the
partial cross sections in the $q \bar q$ channel and $y_0=v_0=0.05$ for the
partial cross sections in the $qg$ and $g\bar q$ channels. For the cut
parameters defined in section V of \cite{MS1} we choose the following values:
$\Delta_x = 10^{-5}$ and $\Delta_y = \Delta_v = 10^{-8}$. As discussed in the
referred section these cuts introduce errors in the numerical results, however,
we have checked that the above choice minimizes these errors while keeping the
Monte Carlo program stable. Our tables of results for total cross sections
include entries for the

lowest order estimates of these errors and confirm that they are negligible.
Thus the expressions for the errors are neglected in all the distributions and
correlations.
\\

{\sc 2) Exclusive production of $W^+$, $\gamma$ and $1$ jet.\\}

In this scenario (``$1$ jet scenario'') one detects three outgoing particles,
namely the $W^+$, $\gamma$ and $1$ jet. This scenario is thus defined by the
following kinematic conditions:
\begin{eqnarray}
\label{1.2}
|\cos\theta_{\gamma}|, |\cos\theta_{W}|,|\cos\theta_{jet}|  &<& \cos(0.3\ {\rm
rad}) \nonumber \\
Pt_{\gamma}, Pt_{W}, Pt_{jet}  &>& 10.0\ {\rm GeV} \nonumber \\
R_{W,\gamma} &>& 0.2 \nonumber \\
R_{jet,\gamma} &>& 0.2 \nonumber \\
R_{jet,W} &>& 0.2 \,.
\end{eqnarray}
The above conditions will automatically remove bremsstrahlung contributions
which tend to produce jets parallel to the outgoing photon. In the Monte Carlo
runs we choose the values for the artificial parameters of this scenario in all
channels as: $x_0 = 1$, $y_0 = v_0 =0$. This choice and the above cuts
guarantee that no unphysical dependence is introduced in the calculation of
physical quantities related to the jet. Moreover, the cuts
$\Delta_x,\Delta_y,\Delta_v$ introduced in the first scenario are not necessary
here and thus the associated errors are zero in this case.
\\

{\sc 3) Exclusive production of $W^+$ and $\gamma$ with $0$ jet.\\}

In this scenario (``$0$ jet scenario'') we select events where the $W^+$ boson
and the photon are detected and no outgoing jet is detected. This will include
$2$ to $3$ body events where the outgoing jet is not detected because it has
either a small angle with respect to the beam or a small transverse momentum or
it is ``shadowed'' by the $W^+$ boson or the photon. The following kinematic
conditions
define this scenario:
\begin{eqnarray}
\label{1.3}
|\cos\theta_{\gamma}|, |\cos\theta_{W}| &<& \cos(0.3\ {\rm rad}) \nonumber \\
Pt_{\gamma}, Pt_{W}  &>& 10.0\ {\rm GeV} \nonumber \\
R_{W,\gamma} &>& 0.2 \nonumber \\
( R_{jet,\gamma} < 0.2 )&\Longrightarrow& (s_{(jet,\gamma)} < 0.2) \nonumber \\
( R_{jet,W} < 0.2 )&\Longrightarrow& (s_{(jet,W)} < 0.2) \nonumber \\
( |\cos\theta_{jet}| > \cos(0.3\ {\rm rad}) )& \rm{or} & (Pt_{jet} < 10.0\ {\rm
GeV}) \,.
\end{eqnarray}
As already discussed in \cite{MS1} the results for this scenario can be
obtained by subtracting the results for the $1$ jet scenario from the
corresponding results for the $2$ body inclusive scenario.
\\

In section II we present results for total cross sections. Section III is
devoted to the discussion of single differential distributions and
correlations. In section IV we present and discuss results for double
differential cross sections and correlations. In section V we end our study
with some concluding remarks.

\newpage
\pagestyle{myheadings}
\mysection{\bf TOTAL CROSS SECTIONS.}
\label{II}

\subsection{\sc Brief Review of Relevant Formulae.}

We write the total hadronic cross section as:
\begin{eqnarray}
\label{2.1}
\sigma^H &=& \int_{0}^{1}d\tau_1\int_{0}^{1}d\tau_2 \left\{
f_{qp}(\tau_1)f_{\bar q\bar p}(\tau_2)\sigma_{q\bar q} + f_{qp}(\tau_1)f_{g\bar
p}(\tau_2)\sigma_{qg} + f_{gp}(\tau_1)f_{\bar q\bar p}(\tau_2)\sigma_{g\bar q}
\right. \nonumber \\
& & \hspace{0.9in} + \left. (p\leftrightarrow \bar p,\tau_1\leftrightarrow
\tau_2,p_1\leftrightarrow p_2,P_1\leftrightarrow P_2) \right\} \nonumber \\
\,
\end{eqnarray}
with
\def\sigtilde{{\tilde\sigma}}
\begin{eqnarray}
\label{2.2}
\sigma_{q\bar q}&=&\sigma_{q\bar q}^{Born} + \sigma^P_{q\bar q\ (SV)} +
\sigma_{Ia} + \sigma_{Ib} + \sigma_{I,4} + \sigma_{q\bar q\ (finite)}^{P} +
\sigma_{q\bar q\ (Brems)}  + \sigma_{q\bar q(error)} \nonumber \\
\sigma_{qg}&=&\sigma_{qg,finite}^{P,I} + \sigma_{qg}^{I,col} +
\sigma_{qg,finite}^{P,II} + \sigma_{qg}^{II,col} + \sigtilde^{P}_{qg} +
\sigma_{qg(Brems)} + \sigma_{qg(error)} \nonumber \\
\sigma_{g\bar q}&=&\sigma_{g\bar q,finite}^{P,I} + \sigma_{g\bar q}^{I,col} +
\sigma_{g\bar q,finite}^{P,II} + \sigma_{g\bar q}^{II,col} +
\sigtilde^{P}_{g\bar q} + \sigma_{g\bar q(Brems)}  + \sigma_{g\bar q(error)}
\nonumber \\
\,
\end{eqnarray}
and
\begin{eqnarray}
\label{2.3}
\sigma_{Ia} &=& \sigma_{Ia,1} + \sigma_{Ia,2} + \sigma_{Ia,3} \nonumber \\
\sigma_{Ib} &=& \sigma_{Ib,1} + \sigma_{Ib,2} + \sigma_{Ib,3} \nonumber \\
\sigma_{q\bar q\ (finite)}^{P} &=& \sigma_{f,1,1,a} + \sigma_{f,1,2,a} +
\sigma_{f,1,3,a} + \sigma_{f,1,1,b} + \sigma_{f,1,2,b} + \sigma_{f,1,3,b}
\nonumber \\
&+&\sigma_{f,2,1,a} + \sigma_{f,2,2,a} + \sigma_{f,2,3,a} + \sigma_{f,2,1,b} +
\sigma_{f,2,2,b} + \sigma_{f,2,3,b} + \sigma_{f,3} \nonumber \\
\sigma_{qg,finite}^{P,I} &=& \sigma_{qg,f,1}^{I} + \sigma_{qg,f,2}^{I} +
\sigma_{qg,f,3}^{I}  \nonumber \\
\sigma_{qg,finite}^{P,II} &=& \sigma_{qg,f,1}^{II} + \sigma_{qg,f,2}^{II} +
\sigma_{qg,f,3}^{II} \nonumber \\
\sigma_{g\bar q,finite}^{P,I} &=& \sigma_{g\bar q,f,1}^{I} + \sigma_{g\bar
q,f,2}^{I} + \sigma_{g\bar q,f,3}^{I} \nonumber \\
\sigma_{g\bar q,finite}^{P,II} &=& \sigma_{g\bar q,f,1}^{II} + \sigma_{g\bar
q,f,2}^{II} + \sigma_{g\bar q,f,3}^{II} \,.
\end{eqnarray}

Each term in (\ref{2.2}) and (\ref{2.3}) has been explained in section V of
\cite{MS1}.
Note that the replacements in parenthesis in (\ref{2.1}) apply to all explicit
as well as implicit variables in the above expressions. The last replacement in
the parenthesis in (\ref{2.1}) will act on physical variables that go into
experimental cut functions and histograms (see section IV.D of \cite{MS1}.)

For the bremsstrahlung contributions we use here the so called leading
logarithm approximation, which has been previously used by other authors
\cite{DO},\cite{O},\cite{O1}. In this approximation we write the photon
fragmentation functions as:
\begin{eqnarray}
\label{2.4}
f_{\gamma q(\bar q)}(x,M) &=&
\frac{\alpha}{2\pi}\ln\left(\frac{M^2}{\Lambda^2_{QCD}}\right)\left[ \frac{
\hat{e}_{q(\bar q)}^2\left( 2.21 - 1.28x +1.29x^2 \right)x^{-0.951} }{
1-1.63\ln(1-x) } \right. \nonumber \\
& &\hspace{1.in}+ 0.002(1-x)^{2.0}x^{-2.54} \bigg]\nonumber \\
f_{\gamma g}(x,M) &=& \frac{\alpha}{2\pi}\ln\left( \frac{M^2}{\Lambda^2_{QCD}}
\right)0.0243(1-x)^{1.03}x^{-1.97}\,,
\end{eqnarray}
where $\hat{e}_{q(\bar q)}$ denotes the charge of the outgoing quark
(antiquark) $q$ ($\bar q$) in units of $e$, $M$ is the mass factorization scale
and the running electromagnetic fine structure constant is
$\alpha=e^2(\mu^2)/4\pi$ .\\

\subsection{\sc Total cross section for the $2$ body inclusive reaction $p +
\bar p \rightarrow W^+ + \gamma + X$.}

All terms in expressions (\ref{2.2}) and (\ref{2.3}) will produce nonvanishing
contributions in this scenario. In table \ref{tab1} we show the results for the
$q\bar q$ channel cross sections in pb at the three scales $r=0.5, 1.0$ and
$2.0$, where $r=M/M_W=\mu/M_W$ and $M,\mu$ are the mass factorization and
renormalization scales, respectively.  The second entry of table \ref{tab1}
shows the value used for $\alpha_S$ at these scales. In all our runs we use
$\alpha = 1/137.036$ for the fine structure constant and $G_F=1.16639\cdot
10^{-5}{\rm GeV}^{-2}$ for the Fermi coupling constant.
\begin{table}
\centerline{
\begin{tabular}{||c|c|c|c||} \hline
$r$                              & $0.50$   &  $1.0$  & $2.0$   \\ \hline\hline
$\alpha_S$                       & 0.145    &  0.129  & 0.116   \\ \hline\hline
$\sigma^{Born}_{q\bar q}$        & 2.66     &  2.62   & 2.58   \\ \hline\hline
$\sigma^{P}_{SV}$                & 0.31     &  0.51   & 0.67   \\ \hline\hline
$\sigma_{Ia,1}$                  & 0.02     &  0.02   & 0.02      \\  \hline
$\sigma_{Ia,2}$                  & 0.05     & -0.05   &-0.11     \\  \hline
$\sigma_{Ia,3}$                  &-18.45    &-18.25   &-17.98     \\
\hline\hline
$\sigma_{Ia}$                    &-18.39    &-18.28   &-18.07      \\
\hline\hline
$\sigma_{Ib,1}$                  & 0.02     &  0.02   &  0.01    \\  \hline
$\sigma_{Ib,2}$                  & 0.04     & -0.04   & -0.10     \\  \hline
$\sigma_{Ib,3}$                  &-18.43    &-18.23   &-17.94      \\
\hline\hline
$\sigma_{Ib}$                    &-18.38    &-18.25   &-18.03      \\
\hline\hline
$\sigma_{I,4}$                   & 37.20    & 36.83   & 36.29      \\
\hline\hline
$\sigma_{f,1,1,a}$               &  0.07    &  0.06   &  0.05     \\  \hline
$\sigma_{f,1,2,a}$               &  1.41    &  1.20   &  1.04    \\  \hline
$\sigma_{f,1,3,a}$               & -1.30    & -1.12   & -0.97     \\  \hline
$\sigma_{f,1,1,b}$               &  0.08    &  0.07   &  0.06     \\  \hline
$\sigma_{f,1,2,b}$               &  1.20    &  1.03   &  0.90    \\  \hline
$\sigma_{f,1,3,b}$               & -1.14    & -0.98   & -0.85     \\  \hline
$\sigma_{f,2,1,a}$               &  1.12    &  0.98   &  0.87    \\  \hline
$\sigma_{f,2,2,a}$               & 28.97    & 26.45   & 23.44     \\  \hline
$\sigma_{f,2,3,a}$               &-29.03    &-26.51   &-23.50      \\  \hline
$\sigma_{f,2,1,b}$               &  1.12    &  0.99   &  0.87    \\  \hline
$\sigma_{f,2,2,b}$               & 29.95    & 26.31   & 23.33     \\  \hline
$\sigma_{f,2,3,b}$               &-30.01    &-26.37   &-23.38      \\  \hline
$\sigma_{f,3}$                   & -2.30    & -2.02   & -1.79     \\
\hline\hline
$\sigma^P_{q\bar q(finite)}$     &  0.16    &  0.10   &  0.07    \\
\hline\hline
$\sigma_{q\bar q(Brems)}$        &  0.003   &  0.003  &  0.003    \\
\hline\hline
$\sigma_{q\bar q(error)}$         &$O(10^{-4})$&$O(10^{-4})$&$O(10^{-4})$\\
\hline\hline
$\sigma_{q\bar q}$               &  3.61    & 3.52    &  3.52    \\
\hline\hline
\end{tabular}
}
\caption{Results for the partial hadronic cross sections (in pb) for the 2-body
inclusive reaction $p + \bar p \rightarrow W^+ + \gamma + X$ in the $q\bar q$
channel at $\protect\sqrt{S}=1.8$ TeV.}
\label{tab1}
\end{table}

The corresponding results for the $qg$ channel contributions to the 2-body
inclusive hadronic
reaction $p + \bar p \rightarrow W^+ + \gamma + X$ are summarized in table
\ref{tab2}.
\begin{table}
\centerline{
\begin{tabular}{||c|c|c|c||} \hline
$r$                              & $0.50$  &  $1.0$  & $2.0$   \\ \hline\hline
$\alpha_S$                       & 0.145   &  0.129  & 0.116   \\ \hline\hline
$\sigma^{I}_{qg,f,1}$            & 0.38    &  0.31   & 0.26    \\  \hline
$\sigma^{I}_{qg,f,2}$            & 0.78    &  0.66   & 0.56    \\  \hline
$\sigma^{I}_{qg,f,3}$            &-0.80    & -0.67   &-0.57     \\
\hline\hline
$\sigma^{P,I}_{qg,finite}$       & 0.38    &  0.31   & 0.26    \\  \hline\hline
$\sigma^{I,col}_{qg}$            &-0.22    & -0.25   &-0.26    \\  \hline\hline
$\sigma^{II}_{qg,f,1}$           & 0.12    &  0.10   & 0.09    \\  \hline
$\sigma^{II}_{qg,f,2}$           & 0.15    &  0.13   & 0.11    \\  \hline
$\sigma^{II}_{qg,f,3}$           &-0.08    & -0.07   &-0.06    \\  \hline\hline
$\sigma^{P,II}_{qg,finite}$      & 0.19    &  0.16   & 0.13    \\  \hline\hline
$\sigma^{II,col}_{qg}$           &-0.04    & -0.04   &-0.04    \\  \hline\hline
$\sigtilde^{P}_{qg}$             & 0.15    &  0.12   & 0.10    \\  \hline\hline
$\sigma_{qg(Brems)}$             & 0.02    &  0.02   & 0.02    \\  \hline\hline
$\sigma_{qg(error)}$             &$O(10^-9)$&$O(10^-9)$&$O(10^-9)$ \\
\hline\hline
$\sigma_{qg}$                    & 0.48    &  0.33   & 0.21    \\  \hline\hline
\end{tabular}
}
\caption{Results for the partial hadronic cross sections (in pb) for the 2-body
inclusive reaction $p + \bar p \rightarrow W^+ + \gamma + X$ in the $qg$
channel at $\protect\sqrt{S}=1.8$ TeV.}
\label{tab2}
\end{table}
Finally, the corresponding results for the  $g\bar q$ channel are summarized in
table \ref{tab3}.
\begin{table}
\centerline{
\begin{tabular}{||c|c|c|c||} \hline
$r$                              & $0.50$   &  $1.0$  & $2.0$   \\ \hline\hline
$\alpha_S$                       & 0.145    &  0.129  & 0.116   \\ \hline\hline
$\sigma^{I}_{g\bar q,f,1}$       & 0.14     &  0.12   & 0.10    \\  \hline
$\sigma^{I}_{g\bar q,f,2}$       & 0.60     &  0.50   & 0.42    \\  \hline
$\sigma^{I}_{g\bar q,f,3}$       &-0.61     & -0.51   &-0.43     \\
\hline\hline
$\sigma^{P,I}_{g\bar q,finite}$  & 0.13     &  0.11   & 0.09    \\
\hline\hline
$\sigma^{I,col}_{g\bar q}$       &-0.18     & -0.20   &-0.20    \\
\hline\hline
$\sigma^{II}_{g\bar q,f,1}$      & 0.38     &  0.31   & 0.26    \\  \hline
$\sigma^{II}_{g\bar q,f,2}$      & 0.48     &  0.40   & 0.34    \\  \hline
$\sigma^{II}_{g\bar q,f,3}$      &-0.27     & -0.23   &-0.20    \\
\hline\hline
$\sigma^{P,II}_{g\bar q,finite}$ & 0.59     &  0.49   & 0.41    \\
\hline\hline
$\sigma^{II,col}_{g\bar q}$      &-0.13     & -0.13   &-0.13    \\
\hline\hline
$\sigtilde^{P}_{g\bar q}$        & 0.11     &  0.09   & 0.07    \\
\hline\hline
$\sigma_{g\bar q(Brems)}$        & 0.08     &  0.08   & 0.07    \\
\hline\hline
$\sigma_{g\bar q(error)}$         & $O(10^{-8})$&$O(10^{-8})$&$O(10^{-8})$\\
\hline\hline
$\sigma_{g\bar q}$               & 0.61     &  0.43   &  0.31    \\
\hline\hline
\end{tabular}
}
\caption{Results for the partial hadronic cross sections (in pb) for the 2-body
inclusive reaction $p + \bar p \rightarrow W^+ + \gamma + X$ in the $g\bar q$
channel at  $\protect\sqrt{S}=1.8$ TeV.}
\label{tab3}
\end{table}

We thus obtain total cross sections for the 2-body inclusive reaction $p + \bar
p \rightarrow W^+ + \gamma + X$ of $4.70,4.28$ and $4.04$ pb for $r=0.5, 1.0$
and $2.0$ respectively. The total cross section at order $\alpha_S$ is thus a
decreasing function of the scale $r$, confirming the results obtained in a
previous study \cite{MSN}. The strong dependence of the cross section on the
scale $r$ means that order $\alpha^2_S$ corrections are large and they must be
included for accurate predictions in this scenario. This was also pointed out
in \cite{MSN} when studying the single particle photon inclusive cross section.

If we compare these results with the ones previously reported for the single
particle inclusive total cross section in \cite{MSN} we note a considerable
reduction of the total cross section in the present calculation. The smaller
values can be attributed to larger cut in the photon angle and the effect of
experimental cuts on the $W^+$ boson which was treated previously in an
inclusive fashion.

We also note that in the approximation we are using here the bremsstrahlung
contributions are negligible in the $q\bar q$ channel and that their
contribution in the other channels amounts to not more than 2\% of the total
cross section. This would make big deviations from the leading logarithm
bremsstrahlung approximation easy to study.

The next to last rows in tables \ref{tab1},\ref{tab2} and \ref{tab3} show that
the errors introduced by the $\Delta_x, \Delta_y$ and $\Delta_v$ cuts (see
section V of \cite{MS1},) are negligible and thus our prescription for
splitting the $x, y$ and $v$ integrals when adding up the histograms is
consistent.

\subsection{\sc Total cross sections of $p + \bar p \rightarrow W^+ + \gamma +
jet$ and $p + \bar p \rightarrow W^+ + \gamma$. }

With the choice of parameters $x_0,y_0$ and $v_0$ explained in section I the
only terms in (\ref{2.2}) contributing to the reaction $p + \bar p \rightarrow
W^+ + \gamma + jet$ are:
\begin{eqnarray}
\label{2.5}
\sigma_{q\bar q} &=& \sigma_{f,1,1,a} + \sigma_{f,1,1,b}  \nonumber \\
\sigma_{qg} &=&\sigma_{qg,f,1}^{I} + \sigma_{qg,f,1}^{II} + \sigtilde^{P}_{qg}
\nonumber \\
\sigma_{g\bar q} &=&\sigma_{g\bar q,f,1}^{I} + \sigma_{g\bar q,f,1}^{II} +
\sigtilde^{P}_{g\bar q} \,.
\end{eqnarray}
The results for the contributions of all these terms to the hadronic cross
section are shown in table \ref{tab4} for the three scales $r=0.5, 1.0$ and
$2.0$.
At these scales we thus obtain for the total cross sections of the $1$ jet
reaction values of $2.15,1.79$ and $1.51$ pb respectively. As in the $2$ body
inclusive case the variation is large in the $1$ jet scenario so we need to
include higher order QCD corrections to make more accurate predictions.
Unfortunately these higher order corrections are not available.
\begin{table}
\centerline{
\begin{tabular}{||c|c|c|c||} \hline
$r$                              & 0.50    &  1.0    & 2.0   \\ \hline\hline
$\alpha_S$                       & 0.145   &  0.129  & 0.116   \\ \hline\hline
$\sigma_{f,1,1,a}$               & 0.54    &  0.46   & 0.39     \\  \hline
$\sigma_{f,1,1,b}$               & 0.52    &  0.44   & 0.38      \\
\hline\hline
$\sigma_{q\bar q}$               & 1.05    &  0.89   & 0.77     \\
\hline\hline
$\sigma^{I}_{qg,f,1}$            & 0.26    &  0.21   & 0.17    \\  \hline
$\sigma^{II}_{qg,f,1}$           & 0.14    &  0.11   & 0.09    \\  \hline
$\sigtilde^{P}_{qg}$             & 0.12    &  0.10   & 0.08    \\  \hline\hline
$\sigma_{qg}$                    & 0.52    &  0.42   & 0.35    \\  \hline\hline
$\sigma^{I}_{g\bar q,f,1}$       & 0.07    &  0.06   & 0.05    \\  \hline
$\sigma^{II}_{g\bar q,f,1}$      & 0.41    &  0.33   & 0.27    \\  \hline
$\sigtilde^{P}_{g\bar q}$        & 0.10    &  0.08   & 0.06    \\  \hline\hline
$\sigma_{g\bar q}$               & 0.58    &  0.47   & 0.39    \\  \hline\hline
$\sigma_{total}$                 & 2.15    &  1.79   & 1.51    \\  \hline\hline
\end{tabular}
}
\caption{Results for the partial and total hadronic cross sections (in pb) for
the reaction $p + \bar p \rightarrow W^+ + \gamma + jet$ at $\protect\sqrt
S=1.8$ TeV.}
\label{tab4}
\end{table}

Subtracting the above numbers from the ones for the $2$ body inclusive reaction
$p + \bar p \rightarrow W^+ + \gamma + X$ we obtain for the $0$ jet reaction $p
+ \bar p \rightarrow W^+ + \gamma$ the values of $2.55,2.50$ and $2.53$ pb at
the scales $r=0.5,1.0$ and $2.0$ respectively. In this scenario we do not need
to include order $\alpha^2_S$ QCD corrections to make our predictions more
reliable, since they are already very stable.

\newpage
\pagestyle{myheadings}
\mysection{\bf SINGLE DIFFERENTIAL CROSS SECTIONS AND CORRELATIONS.}
\label{III}

We now turn to the analysis of the single differential distributions and
correlations. Our results are shown in figures 1 to 25. The error bars
represent the theoretical uncertainty associated with the dependence of our
results on the mass factorization and renormalization scales. They have been
obtained evaluating the distributions at the two scales $r=0.5$ and $2.0$. Note
that the central values and the upper and lower limits still contain a
numerical error introduced via the Monte Carlo. This error is negligible in our
$1$ jet results but it is larger in our $2$ body inclusive and $0$ jet results.
Regions with big error bars in our plots may be interpreted as regions where
perturbation theory at order $\alpha_S$ is not reliable and thus higher order
QCD corrections would be needed to make more accurate theoretical predictions.

In general, the single differential distributions and correlations have very
little dependence on the scale $r$ in the scenario where the $W^+$ and photon
are produced with $0$ outgoing jet. If we include contributions from $1$ jet
processes we increase the statistics by more than 60\%, but a nonnegligible
theoretical uncertainty is introduced by the scale dependence of the $1$ jet
processes. To make theoretical predictions for single distributions and
correlations more reliable in the $2$ body inclusive scenario, we would thus
need to include the order $\alpha^2_S$ corrections.

Figures 1 to 6 show results for single differential distributions for the $W^+$
boson and figures 7 to 11 show the corresponding photon distributions. These
plots demonstrate that the $W$ boson is mainly concentrated in the regions of
small energy and transverse momentum in all three scenarios. In the $1$ jet
scenario $E_W$ peaks at around $85$ GeV (figure 1) while $Pt_W$ peaks at around
$30$ GeV (figure 2). In the other two scenarios the single differential
distributions of $E_W$ and $Pt_W$ are monotonically decreasing functions except
for the physical and experimental lower cuts.

Figures 3, 4 and 6 show clearly that there are dips in the polar angle and
pseudorapidity distributions of the $W$ boson at around $\theta_W = \pi/2$ in
the $0$ jet and $2$ body inclusive scenarios, with smooth dips at around the
same point in the $1$ jet scenario. This is very different to what we observe
for the photon polar angle and rapidity distributions, which present dips at
around $\theta_W = 2\pi/3$ in the $0$ jet scenario only (figures 9, 10 and 11).

The $0$ jet scenario would so far be the best for studying deviations from the
Standard Model by looking at photon distributions since on the one hand the
theoretical uncertainties related to dependences on the scale $r$ are small and
on the other hand the $2$ to $3$ body contributions are suppressed so that the
partonic zeros still show up as wide dips in the angular and rapidity
distributions of the photon (see figures 9, 10 and 11). The $\theta_{\gamma}$
distribution maintains its features when $1$ jet events are added, as seen in
figure 9, however, the dip in rapidity is smeared out, as seen in figure 11. If
photon bremsstrahlung contributions are not well accounted for by the leading
logarithm approximation, these may disturb the dips in the photon angle and the
photon rapidity and all other photon single and double differential
distributions and correlations in the $0$ jet scenario. Therefore, to isolate
this effect it may be important to look also at physical quantities in the $1$
jet scenario, where

there is no contribution from photon bremsstrahlung at present order in
perturbation theory.

Figures 12 to 15 show single differential distributions of the outgoing jet in
the $1$ jet scenario. We see that the jet is concentrated in the low energy,
low $Pt$ and central rapidity regions. We also note that the scale dependence
increases in these regions. The plot in figure 14 shows a quite uniform polar
angle distribution for the jet.

Figures 16 to 25 summarize our results for single correlations between outgoing
pairs of particles. The same comments about the scale dependences in the
different scenarios observed in figures 1 to 15 are valid here.

The $R_{W,\gamma}$ cone size correlation in figure 16 presents a sharp peak
followed by a sharp dip that falls below zero near $R_{W,\gamma} = \pi$ in the
$0$ jet and $2$ body inclusive scenarios. A similar feature was observed by
Mangano et al. \cite{MNR} in $R_{q,\bar q}$ for heavy quark pair correlations
with fixed quark rapidity. The corresponding correlation in the $1$ jet process
presents no anomalous behavior, suggesting that the observed dips in the other
two scenarios are the effect of the subtraction pieces introduced when the
factorization theorem is implemented to cancel collinear singularities
associated with untagged jets produced parallel to the beams.

The $R_{W,jet}$ cone size correlation in figure 21 has a very symmetric shape
with a peak centered at $\pi$. On the other hand, $R_{jet,\gamma}$ in figure 21
is much smoother between $0.2$ and $\pi$, falling rapidly to zero outside this
range. This means that the jet and the photon are approximately uncorrelated in
cone size, however the jet and the $W$ boson are highly correlated. Figure 25
confirms that the latter pair of particles are mainly concentrated on a plane
that contains the beams, i.e. the distribution of the azimuthal angle
difference $\phi_{W,jet}$ peaks at $\pi$. However, $\phi_{jet,\gamma}$ is
approximately flat over the whole range. The observed bump near
$\phi_{jet,\gamma} = R_{jet,\gamma} = 0.2$ is caused by the required
experimental cut on the cone size.

Pair mass correlations are of interest when studying deviations from the
Standard Model, as pointed out in \cite{O1}. The pair mass correlations
$M_{W,j} = \sqrt{ (Q_1 + p_j)^2}$ with $j=\gamma$, jet in figures 17 and 22
both peak at around $100$ GeV, however the correlation $M_{jet,\gamma}$ has a
plateau between $20$ and $40$ GeV slowly falling to zero above $40$ GeV.

Figures 18 and 23 show the distributions of the angles $\alpha_{i,j}$ between
pairs of particles $i,j$. As expected $\alpha_{W,\gamma}$ peaks near $\alpha =
\pi$ in the $0$ jet scenario, since here the main contributions come from back
to back $2$ to $2$ body partonic reactions. The smearing is introduced when
boosting the partonic system into the hadronic system. This is not the case for
the corresponding correlation in the $1$ jet scenario, which shows that the $W$
boson and the photon are uncorrelated in the angle difference. The angular
correlations between the jet and the $W$ or the photon are also quite smooth.

Since transverse momenta are preserved under Lorentz boosts along the beamline,
the distribution of the azimuthal angle difference $\phi_{W,\gamma}$ only makes
sense in the $1$ jet scenario, which is shown in figure 19. It also peaks at
around $\phi_{W,\gamma} = \pi$, meaning that the $W$ and the photon are mainly
concentrated on the plane that contains the beams. This, along with the
information contained in figure 25, shows that the beams and the three
particles produced in the $1$ jet scenario are approximately coplanar.

Figure 20 shows the distributions of the pseudorapidity difference between the
$W$ and the photon in the three scenarios. In all cases the events are
concentrated along the negative region, which means that the photon is produced
with an angle with respect to the incoming antiproton beam that is usually
smaller than the corresponding angle of the $W$ boson. We note the presence of
two dips in the $0$ jet scenario: one around $\eta^{\ast}_{W,\gamma} = -1$ and
the other around $\eta^{\ast}_{W,\gamma} = 1.3$. These dips are smeared into
plateaus in the inclusive scenario by the effect of $1$ jet processes. In
figure 24 we see a very different behavior for the pseudorapidity differences
of jet and $W$ or photon: $\eta^{\ast}_{W,jet}$ is symmetric around $0$ while
$\eta^{\ast}_{jet,\gamma}$ receives bigger contributions on the negative axis
but both correlations peak at $0$.

\newpage
\pagestyle{myheadings}
\mysection{\bf DOUBLE DIFFERENTIAL CROSS SECTIONS AND CORRELATIONS.}
\label{IV}

Double differential cross sections and correlations help us visualize the
qualitative features of the spatial distributions of particles hitting an
experimental detector. We present in figures 26.a to 45 results obtained by
averaging Monte Carlo runs at $r=0.5$ and $r=2.0$. Figures 26.a to 27.c
correspond to double differential distributions of the $W$ while figures 28.a
to 29.c show the corresponding double differential distributions of the photon.
Figures 30.a to 34.c contain the plots of the correlations between the $W$
boson and the photon and figures 35 to 45 the various correlations between the
jet and either the $W$ or the photon.

First we look at the double differential distributions of $Pt$ vs $\theta$ for
the $W$ boson (figures 26.a, b and c). We conclude that the dip in $\theta_W$,
already observed in the single differential distribution (figure 3), increases
in relative depth as $Pt_W$ decreases. In the $1$ jet scenario the double
differential distribution clearly peaks along the line $Pt_W = 30$ GeV for all
values of $\theta_W$ (figure 26.b). In the other two scenarios the peaks move
down to small values of $Pt_W$, but in figures 26.a and 26.c they are not
clearly seen due to the low density of bins that have been used on each axis;
these peaks show up in the single differential distributions in figure 2,
however, where the density of bins is four times bigger. Similar conclusions
are obtained when looking at $Pt$ vs pseudorapidity for the $W$ in figures
27.a,b and c.

The corresponding double differential distributions for the photon have been
plotted in figures 28.a to 29.c. The effect of the $1$ jet processes on the
dips in angle and rapidity which reflect the partonic zeros is clearly visible
in these figures. We note that in the $0$ jet scenario the dips increase their
relative depth as $Pt_{\gamma}$ decreases. This suggests that the effects of
deviations from the Standard Model on the single differential distributions in
the polar angle and rapidity of photon would be enhanced if $Pt_{\gamma}$ is
constrained to take values below certain bound, say $25$ GeV.

{}From the results in the previous subsection we already know that the events
are mainly concentrated in the regions of small $E_W$ and small $E_{\gamma}$.
Moreover, figures 30.a, b and c show that
$E_W$ and $E_{\gamma}$ are not very correlated and that the bulk of the events
is always in the small $E_W(\gamma)$ region, regardless of the value of
$E_{\gamma}(W)$. The peak in the double differential cross section is steepest
in the $0$ jet scenario, however (see figure 30.c).

Since $Pt$ is conserved under Lorentz boosts along the beams, the $W$ and the
photon are exactly correlated in transverse momentum space in the $0$ jet
scenario (i.e. $Pt_W = Pt_{\gamma}$), so this correlation only makes sense in
the $1$ jet scenario, where it gets smeared and it peaks around the line $Pt_W
= 30 GeV$ regardless of the value of $Pt_{\gamma}$ (see figure 31).

The pseudorapidity correlation between $W$ and $\gamma$ clearly shows the
effect of the partonic zero as we can appreciate in figure 32.c in the $0$ jet
scenario. The double differential cross section shows a valley in the negative
$\eta_{\gamma}$ region along the whole range of $\eta^{\ast}_{W}$. Inclusion of
$1$ jet processes (see figure 32.b) would push the valley up producing a
plateau instead, as seen in figure 32.a for the $2$ body inclusive scenario and
smearing out the dip in the single distribution, as already noted in figure 11.

Due to the exact correlation between $Pt_W$ and $Pt_{\gamma}$ in the $0$ jet
scenario, the correlation between
$Pt_{W}$ and $\eta_{\gamma}$ in the $0$ jet scenario (figure 33.c) looks
similar to the double differential distribution of $Pt_{\gamma}$ vs
$\eta_{\gamma}$ (figure 29.c), i.e., it shows an increasing effect of the
partonic zero as $Pt_{W}$ decreases so that the photon rapidity dip becomes
relatively deeper at small $Pt_{W}$. $1$ jet processes present no dip in this
double correlation, as seen in figure 33.b, and their effect when looking at
the $2$ body inclusive process in figure 33.a is to smear out the dip, which
reappears below $Pt_{W}=30$ GeV.

The pseudorapidity of the $W$ shows dips in all three scenarios (figure 6) and
these dips are enhanced if $Pt_{\gamma}$ is constrained to be smaller than $30$
GeV, as inferred from figures 34.a, b and c. We contrast this with the results
for the rapidity of the photon, which only shows a clear dip in the $0$ jet
scenario, even when looking at regions of low $Pt_{\gamma}$ (figures 29.a, b
and c).

The $Pt_{jet}$ correlations with $Pt_{\gamma}$ and $Pt_W$ are quite different:
in the $Pt_{\gamma}$ versus $Pt_{jet}$ space the events are mainly concentrated
along the small $Pt_{\gamma}$ and small $Pt_{jet}$ regions (see figure 35)
while in the $Pt_{W}$ versus $Pt_{jet}$ space they are concentrated along the
$Pt_{W} = Pt_{jet}$ line (see figure 36). Something similar happens in the case
of $\eta_{jet}$ correlations with $Pt_{\gamma}$ and $Pt_W$: in the
$Pt_{\gamma}$ versus $\eta_{jet}$ space the events are concentrated mainly
along the small $Pt_{\gamma}$ regions (see figure 42) while in the $Pt_{W}$
versus $\eta_{jet}$ space they peak along the $Pt_{W} = 30$ GeV line (see
figures 41 and 2).

Although the single rapidity distribution of the photon does not present a dip
in the $1$ jet scenario (figure 11), the dip in $\eta_{\gamma}$ reappears if we
fix the rapidity of the jet, as can be appreciated in figure 37. The dip moves
up in $\eta_{\gamma}$ as $\eta_{jet}$ is increased and it is steeper in the
regions of negative $\eta_{jet}$. This figure also shows that the double
differential cross section peaks along the line $\eta_{\gamma} = \eta_{jet}$.

Figures 43 to 45 show the correlations between cone sizes $R_{i,j}$.
$R_{jet,\gamma}$ versus $R_{W,jet}$ (figure 43) shows a symmetric shape around
the plane $R_{W,jet}=\pi$ at which the correlation peaks with a dip around
$R_{jet,\gamma}=1$; these features still show up in the single distributions in
figure 21. The correlation of $R_{W,jet}$ with $R_{W,\gamma}$ (figure 45) peaks
around $R_{W,jet}=R_{W,\gamma}=\pi$ and also has a symmetric shape with respect
to the plane $R_{W,jet} = \pi$, confirming again the observations regarding the
plot of $R_{W,jet}$ in figure 21. Although $R_{jet,\gamma}$ versus $R_{W,jet}$
(figure 44) shows no simple symmetry it also peaks near the planes
$R_{jet,\gamma} = \pi$ and $R_{W,jet} = \pi$. In general the $R_{i,j}$ cone
sizes peak around $R_{i,j} = \pi$ and fall off quickly above $R_{i,j} = 4$,
and, in particular, $R_{W,jet}$ is symmetric around $\pi$ even when plotted
against other variables.

\newpage
\pagestyle{myheadings}
\mysection{\bf CONCLUDING REMARKS.}
\label{V}
\bigskip

We have completed an analysis of the exclusive reactions $p \bar p \rightarrow
W^+ \gamma$ and $p \bar p \rightarrow W^+ \gamma + jet$ by generalizing the
methodology used by other authors in the context of heavy quark and $Z$ pair
production \cite{MNR},\cite{MNR2}, which consistently includes all divergent
regions of phase space in the framework of the factorization theorem and
dimensional regularization. We have taken into account order $\alpha_S$ QCD
corrections and leading logarithm bremsstrahlung contributions and used the
latest CTEQ set of parton distribution functions, which fit the newly released
HERA data. We have treated the $W$ boson as a real particle with a mass of
$80.2$ GeV.

All our analytical results were presented in our paper \cite{MS1} and have been
used in a Fortran code with Monte Carlo integration techniques. Histograms of
single and double differential cross sections and correlations for all outgoing
particles ($W$ boson, photon and, when applicable, jet) have been obtained in
each of three experimental scenarios, namely $2$ body inclusive, $1$ jet and
$0$ jet scenarios.

Our results for total, single and double differential cross sections and
correlations for the production of $W^+$ and photon accompanied by $0$ jet show
the smallest theoretical uncertainty under variations of the mass factorization
and renormalization scales when compared with the predictions in the $2$ body
inclusive and $1$ jet scenarios. This means that accurate predictions are
available in the $0$ jet scenario for quantities related to the $W$ boson and
the photon without including order $\alpha^2_S$ corrections, however, they
would have to be included for better accuracy in the other two scenarios.

Previous work on the reaction $p \bar p \rightarrow W^+ \gamma +X$ has been
devoted to the study of single photon distributions, photon-$W$ boson pair mass
and photon-charged lepton pseudorapidity correlations. The complete set of
distributions and correlations including the $W$ boson and the jet therefore
complements the studies of the Electroweak sector of the Standard Model (i.e.
the magnetic moment of the $W$) and provides further checks for the QCD sector
and the photon bremsstrahlung process.
We note here that our results for the $0$ jet and $2$ body inclusive reactions
include leading logarithm photon fragmentation functions, so deviations of the
observed experimental data from theoretical predictions could in part be
accounted for by errors introduced by this approximation.
To discriminate the effect of poorly known photon bremsstrahlung contributions
from the effects of deviations from the Standard Model, the analysis of
distributions and correlations in the $1$ jet scenario, which, at the present
order in perturbation theory are free of photon bremsstrahlung, would provide a
valuable tool.
With regard to total cross sections the two reactions  $p \bar p \rightarrow
W^+ \gamma$ and $p \bar p \rightarrow W^+ \gamma + jet$ are of similar
importance, but, as we pointed out before, the latter reaction really requires
even higher order QCD corrections to provide more accurate predictions.

\bigskip

\centerline {\bf Acknowledgements}
\bigskip

S.M. would like to thank Prof.\ D.\ Soper for some clarifying discussions
during the CTEQ '93 summer school. The work in this paper was supported in part
by the contract NSF 9309888.

\newpage
\pagestyle{myheadings}
%

\end{document}